# CARVE: Practical Security-Focused Software Debloating Using Simple Feature Set Mappings

Michael D. Brown, *Georgia Institute of Technology*   Santosh Pande, *Georgia Institute of Technology*

**Abstract**

Software debloating is an emerging field of study aimed at improving the security and performance of software by removing excess library code and features that are not needed by the end user (called bloat). Software bloat is pervasive, and several debloating techniques have been proposed to address this problem. While these techniques are effective at reducing bloat, they are not practical for the average user, risk creating unsound programs and introducing vulnerabilities, and are not well suited for debloating complex software such as network protocol implementations.

In this paper, we propose CARVE, a simple yet effective security-focused debloating technique that overcomes these limitations. CARVE employs static source code annotation to map software features source code, eliminating the need for advanced software analysis during debloating and reducing the overall level of technical sophistication required by the user. CARVE surpasses existing techniques by introducing *debloating with replacement*, a technique capable of preserving software interoperability and mitigating the risk of creating an unsound program or introducing a vulnerability.

We evaluate CARVE in 12 debloating scenarios and demonstrate security and performance improvements that meet or exceed those of existing techniques.

## 1. Introduction

Software bloat affects almost all software, negatively impacting security and performance in a variety of ways [1, 6, 7]. The primary source of bloat is feature creep, the tendency of software packages to accumulate new features over time [8]. End users are unlikely to use all features provided by a software package; unnecessary features are a security risk as they may contain software vulnerabilities or be useful to an attacker crafting a code reuse exploit. Another source of bloat is a result of software engineering practices that favor re-use of modular software libraries. Programs typically use a small subset of the functionality provided by each external library [1], yet the entire library is loaded into the program's memory space at runtime. As is the case with unnecessary features, excess library code may be useful for crafting code reuse attacks and also increases memory consumption at runtime.

Several debloating techniques [2-5, 9-10, 18] have been proposed that promise to improve software security and performance by eliminating unnecessary features and excess library code. While these approaches have been shown to be effective at reducing bloat, they have serious limitations that diminish their potential for use in real-world software engineering projects and security-focused debloating.

### 1.1. Motivation

Existing feature-based debloating techniques rely on complex specifications and advanced software analysis techniques. They have a high technical barrier to entry, and are outside the reach of the typical user. For example, CHISEL [3] requires a user-provided test script that compiles the source code, executes it with inputs corresponding to desired and undesired features, and tests the output of each run to determine if the resulting program is correct [14]. Even for small programs, CHISEL requires long, complex scripts that test a large number of concrete inputs to produce debloated variants. Other approaches, such as TOSS and TRIMMER [2, 4], require the user to interact with sophisticated tools such as LLVM, angr, and TEMU [11-13] to debloat software.

Despite their complexity, existing techniques are not capable of preserving desirable software properties such as graceful error handing, user experience, specification compliance, and interoperability. Because these techniques are only capable of removing code, their resulting debloated programs cannot respond in a meaningful way to attempts to invoke debloated functionality. For example, invoking debloated behavior in CHISEL benchmark programs [14] results in the process terminating with no meaningful output. Network protocol implementations are particularly affected in terms of interoperability. Specification-compliant requests for removed functions made to server software debloated by TOSS cannot respond in a protocol compliant manner, as there is no longer code to handle such requests. The client will likely misinterpret this behavior as a connection failure rather than an unimplemented feature, for which there is typically a well-formed response (HTTP code 501, FTP code 502, etc.).

Existing approaches also have several limitations with respect to debloating for security. Techniques that use dynamic analysis to guide debloating such as CHISEL and TOSS can produce unsound programs and introduce new vulnerabilities as a result of the imprecise manner in which feature related code is identified [18]. New vulnerabilities can also be introduced by debloaters operating at the source code level that are only capable of code removal (e.g. CHISEL), due to the potential for interaction between feature-specific code and code not associated with a feature. As shown by Brown and Pande [22], techniques that permanently alter the program's representation can introduce new code reuse gadgets in unpredictable ways. Gadget introduction can negatively impact security, and properly mitigating this effect may require multiple debloating/analysis iterations to achieve a desirable result. Techniques such as CHISEL are not well suited for this task, as debloating operations can take hours to complete [3].

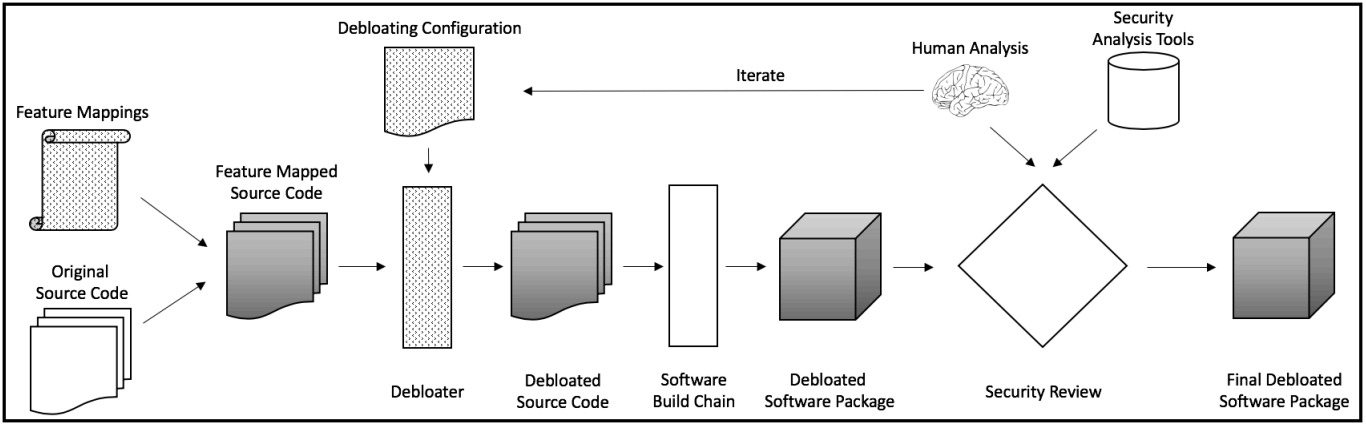

Figure 1: Overview of CARVE.

**1.2. Contributions**

In this paper, we present CARVE: a simple yet effective security-focused debloating technique designed to overcome the limitations of existing approaches. CARVE (Figure 1) does not use dynamic analysis to identify feature related code; instead, users statically map features directly to the source code that implements them using simple and flexible comment-based mappings similar to those commonly used by the software engineering community. These feature mappings allow the user to perform sound, fine-grained debloating that optionally replaces debloated code with replacement code that preserves high level program properties like protocol compliance and prevents the introduction of vulnerabilities. CARVE debloats software in a preprocessing pass, taking feature mapped source code and a configuration file identifying undesired features as input. CARVE scans the source code and intelligently removes code associated with the undesired features, inserting replacement code as indicated by the mappings. Debloating is fast and efficient, as it is decoupled from feature mapping. The final product is a reduced version of the source code containing only desired features that can processed without making changes to the build chain. We discuss the implementation details of CARVE in Section 2.

We evaluated CARVE by using it to debloat features from four network protocol implementations at varying levels of aggressiveness. Debloating this class of software is challenging due to its input / output complexity, compliance to specification, interoperability across platforms, and stateful nature. We analyzed the debloated variants produced by our approach and observed security and performance benefits including vulnerability elimination, reduction in code reuse gadget set utility, code size reduction, and external dependency elimination. We present the results of our evaluation in Section 3. We compare CARVE to existing techniques, discuss its limitations, and identify future work in Section 4.

## 2. Design

Feature-based debloating techniques must perform two tasks: mapping software features to the source code that implements them and removing feature code according to a user specification. Existing techniques [2, 3] approach the first task dynamically by executing the program with concrete inputs corresponding to desired and undesired features, and analyzing the resulting execution paths or program output. These methods have several limitations when used for debloating. First, they require the user to provide a complex specification or use advanced software analysis tools to generate this mapping. They also cannot be used to identify software features that are non-deterministic, triggered by external conditions (e.g. congestion control), or do not produce observable output (e.g. performance features). Scalability is also a concern, since a large number of executions may be required to ensure that the generated mapping is correct. As a consequence, the cost of these approaches grows with the input space and execution time of the target program, which may prove cost-prohibitive for large, feature-rich programs.

CARVE overcomes these limitations by using a static feature mapping technique in which the user places mappings directly in the target program's source code. Embedding the feature mapping as comment-based metadata is familiar task for software developers; this method is used in popular automated software documentation tools such as Doxygen and Sphinx [24, 25] and functions similarly to C/C++ preprocessor directives. Additionally, comment-based feature mappings are not limited to input-triggered features, and can be placed at instruction-level granularity.

The primary disadvantage to this approach is the manual effort required to annotate the source code with feature mappings. Depending on the size of the code base and complexity of the features to be marked as debloatable, the manual effort required to generate feature mappings may be significant. However, we found that this was not the case in practice. Modern software engineering places heavy emphasis on code modularity, refactoring, and separation of concerns, which allows for one to one mapping of the majority of software features to their associated source code.

CARVE further mitigates this disadvantage by incorporating several design elements that reduce the effort required for the user to produce the initial set of feature mappings. Since feature mappings are static and persist within source code, the cost to produce the initial mapping need only be paid once. Additionally, software features can be nested within a feature hierarchy, simplifying the task of mapping a segment of code to multiple related features. CARVE also supports implicit feature mappings that trigger syntax-aware debloating for common code constructs, which automatically handle control-flow implications of code removal. Implicit mappings provide rich debloating support that goes far beyond that of preprocessor directive schemes.

### 2.1. Feature Mapping Anatomy

Figure 2 contains sample code with embedded feature mappings. Feature mappings consist of three parts. The first part is a user configurable tag (`///`), that differentiates code comments from feature mappings. Immediately following the tag, the user specifies one or more feature names, each enclosed within a pair of square braces (`[ ]`). This feature list marks the mapped code as debloatable if the user specifies all of the features for removal. The final part is the optional operator, which is used to differentiate between explicit file mappings, explicit segment mappings, and implicit mappings.

### 2.2. Explicit Feature Mappings

CARVE supports two types of explicit feature mappings (**bold blue** in Figure 2). File mappings are indicated by the `!` operator, and instruct the debloater to remove all code in the file if the specified features are all undesired. Segment explicit mappings are indicated by the `~` operator, and instruct the debloater to remove code between the mapping and the next occurring termination tag (`///~`). Replacement code segments can also be specified between the two replacement tags (`///^`) for segment for segment explicit mappings. For example, the debloating with replacement mapping on line 13 instructs the debloater to remove the code on lines 17-19, and replace it with the code on line 15.

### 2.3. Implicit Feature Mappings

Implicit feature mappings (**bold green** in Figure 2) reduce the effort required to generate feature mappings by offloading two key tasks to the debloater. First, implicit mappings do not require a termination tag to mark the end of the code segment associated with a mapping. Implicit mappings instruct the debloater to analyze the code following the mapping to determine its structure, and by extension what code should be removed. Implicit mappings also obviate the need to handle predictable control-flow implications associated with certain types of code constructs by using explicit mappings with replacement code. For example, mappings preceding switch cases or conditional branches instruct the debloater to analyze

**Figure 2: Example Code with Feature Mappings.**

```
0:  ///[FeatureGroup_A][Feature_G]!
1:
2:  #include <stdio.h>
3:
4:  ///[Feature_Y]~
5:  int a = 0;
6:  int b = 1;
7:  ///~
8:
9:  ///[Feature_Z]
10: char msg[10];
11:
12: int func_y (int i, int j){
13:     ///[Feature_Y]~
14:     ///^
15:     ///return 0;
16:     ///^
17:     int ret = i + a;
18:     int temp = b - j;
19:     return ret * temp;
20:     ///~
21: }
22: ///[Feature_Z]
23: void func_z(int k){ printf(msg); }
24:
25: ///[Feature_G]
26: int func_g (int i){
27:     int j = func_y(i, 3);
28:     ///[Feature_Z]
29:     if(j<0){ func_z(j); }
30:     else{ j = j + i; }
31:     return j;
32: }
```

the control-flow implications of removing the mapped code, and debloat the code in way that does produce unintended control-flow for the resulting debloated program. Examples of implicit mappings are shown in Figure 2. When processing the mapping on line 9, the debloater will recognize that the code following the mapping is a single statement, and remove only line 10. The mappings on lines 22 and 25 target function definitions; in these cases, the debloater will scan for the closing brace associated with the function definition to determine the which lines to remove. The debloater will remove line 23 and lines 26-32 for these mappings, respectively.

Figure 3 shows examples of debloating code constructs with control-flow implications. The code on the right results from debloating `Feature_B` from the code on the left. During syntax-aware analysis of the switch case on line 4, the debloater will determine that the mapped code on lines 6 and 7 are reachable due to the absence of a `break` statement. In this case, the debloater cannot remove these lines and can only remove the case label. Similarly, the debloater will analyze the conditional branch structure on lines 12 and 13 and determine that the presence of an `else` block in the branching construct prevents removal of the entirety of line 12. To maintain the correct branching behavior for the `else` block, only the body of the `else if` block can be debloated.

Processing implicit mappings in the debloater requires syntax-aware analysis that is specific to different programming languages. For this reason, CARVE's design supports the

Figure 3: CARVE Transformations for Code with Control-Flow Implications.

| Before Debloating | After Debloating |
|---|---|
| ```
0:  // Switch Case Example
1:  switch (variable) {
2:      ///[Feature_A]
3:      case FEATURE_A_SPECIFIED:
4:      ///[Feature_B]
5:      case FEATURE_B_SPECIFIED:
6:        length = 2;
7:        break;           }
8:  // If / Else if / Else Example
9:  ///[Feature_A]
10: if (cond_A)      { handle_A(input); }
11: ///[Feature_B]
12: else if (cond_B) { handle_B(input); }
13: else             { handle_generic(input); }
``` | ```
0:  // Switch Case Example
1:  switch (variable) {
2:      ///[Feature_A]
3:      case FEATURE_A_SPECIFIED:
4:      /// Case Label Debloated.
5:        length = 2;
6:        break;           }
7:  // If / Else if / Else Example
8:  ///[Feature_A]
9:  if (cond_A)      { handle_A(input); }
10: else if (cond_B) { /// Code Block Debloated.}
11: else             { handle_generic(input); }
``` |

creation of custom, language-specific debloating modules for future expansion. The resource debloater for C/C++ supports implicit mappings for individual statements, function and struct definitions, switch cases, and conditional branches.

### 2.4. Expressive Power

We found that implicit feature mappings are sufficiently expressive to correctly identify code implementing a feature in the vast majority of cases. Due to the flexibility of high-level programming languages, there are situations in which implicit mappings are insufficient. In these cases, segment explicit mappings should be used with replacement code if necessary. These mappings are capable of arbitrary removal and replacement of code, and are sufficiently expressive for any debloating operation. Across all four of our benchmarks, use of explicit mappings in this manner was required for only 5 of 668 total mappings.

### 2.5. Debloater Operation

The static feature mapping scheme used in CARVE decouples the task of mapping features to code from the task of debloating the code. As a result, debloating is a straightforward and inexpensive operation. CARVE takes as input a single configuration file specifying the location of the feature mapped source code, the programming language the code is written in (used to select the correct syntax-aware debloating module), the hierarchy of all debloatable features, and the list of features the user wishes to debloat. When executed, the debloater makes a copy of the specified source code directories, and begins scanning the feature mapped source code. When feature mappings are found, the debloater compares them to the features specified for debloating. If the feature mapping is a subset of or equal to the features specified for debloating, the debloater processes the mapping. Otherwise, the debloater continues the scan until all files have been scanned.

### 2.6. Soundness

Assuming that the user provided feature mapping (including replacement code) is correct, CARVE is a sound debloating technique. The individual transformations performed by the debloater are sound with respect to program syntax and control-flow. The static feature mapping scheme is fine-grained and directly maps features to the source code that implements them; therefore, a correct mapping does not miss code associated with features or mis-associate code with a feature it does not implement. The same cannot be said for dynamic approaches. CHISEL has been shown to introduce vulnerabilities and cause unexpected program behavior in the author's benchmarks due to limitations inherent to dynamic analysis [18]. The feature identification mechanism used by TOSS cannot guarantee that all feature code is identified; post-debloating fuzzing of both desired and undesired functionality is required to test for unsound variants [2].

Debloating source code is challenging due to its highly expressive nature. There are instances where it is not possible to remove code without producing an unsound program or introducing a vulnerability, especially when considering high level security policies. Consider the example in Figure 4. Debloating either case from the switch statement does not violate language semantics, but does result in the execution of the protected function without prior authorizing the credentials. CARVE mitigates this by supporting debloating with replacement, allowing the user to properly handle the implications of removing source code. In our example, the switch case can be replaced during debloating with exception handling code that traps execution before the protected function.

As is the case with writing source code, creating a feature mapping can be an error prone process. In practice, errors made during the creation of a feature mapping can be identified using the same tools used to ensure that the source code itself is sound. Mapping errors that result in violations of language semantics will be caught by the compiler. Mapping errors present in code that successfully compiles can be

Figure 4: Vulnerability Introduction via Code Removal

```
switch(auth_type){
    case 1: perform_auth_1(creds); break;
    case 2: perform_auth_2(creds); break;
}
protected_function();
```

Table 1: Summary of Security Improvements (C: Conservative; M: Moderate; A: Aggressive).
Values Indicate Count (and Number Reduced) of Expressive Classes and Gadgets.

| Debloated Variant | CVEs Removed | Gadget Set Expressivity | | | Special Purpose Gadget Availability | | | | | |
|---|---|---|---|---|---|---|---|---|---|---|
| | | Practical ROP Exploits | ASLR-Proof ROP Exploits | Turing Complete Exploits | Syscall Gadgets | JOP Dispatcher Gadgets | JOP Data Loader Gadgets | JOP Trampoline Gadgets | COP Dispatcher Gadgets | COP Intra Stack Pivot Gadgets |
| libmodbus (C) | 0 of 0 | 6 (0) | 10 (3) | 6 (1) | 0 (0) | 0 (0) | 3 (-2) | 0 (0) | 0 (0) | 0 (0) |
| libmodbus (M) | 0 of 0 | 6 (0) | 13 (0) | 7 (0) | 0 (0) | 0 (0) | 1 (0) | 0 (0) | 0 (0) | 0 (0) |
| libmodbus (A) | 0 of 0 | 6 (0) | 13 (0) | 7 (0) | 0 (0) | 0 (0) | 1 (0) | 0 (0) | 0 (0) | 0 (0) |
| Bftpd (C) | 0 of 0 | 7 (-1) | 12 (3) | 6 (1) | 0 (0) | 0 (0) | 9 (1) | 0 (0) | 0 (0) | 0 (0) |
| Bftpd (M) | 0 of 0 | 7 (-1) | 17 (-2) | 6 (1) | 0 (0) | 0 (0) | 1 (9) | 0 (0) | 0 (0) | 0 (0) |
| Bftpd (A) | 0 of 0 | 6 (0) | 11 (4) | 5 (2) | 0 (0) | 0 (0) | 1 (9) | 0 (0) | 0 (0) | 0 (0) |
| libcurl (C) | 1 of 6 | 9 (0) | 26 (-1) | 11 (-1) | 5 (-1) | 7 (1) | 50 (1) | 0 (1) | 304 (15) | 4 (0) |
| libcurl (M) | 2 of 6 | 9 (0) | 24 (1) | 10 (0) | 0 (4) | 4 (4) | 41 (10) | 0 (1) | 287 (32) | 3 (1) |
| libcurl (A) | 1 of 6 | 10 (-1) | 24 (1) | 10 (0) | 4 (0) | 3 (5) | 14 (37) | 1 (0) | 170 (149) | 2 (2) |
| mongoose (C) | 5 of 17 | 7 (0) | 10 (6) | 8 (0) | 0 (0) | 1 (-1) | 6 (0) | 0 (0) | 0 (0) | 0 (0) |
| mongoose (M) | 1 of 17 | 7 (0) | 10 (6) | 8 (0) | 0 (0) | 0 (0) | 6 (0) | 0 (0) | 0 (0) | 0 (0) |
| mongoose (A) | 8 of 17 | 7 (0) | 10 (6) | 8 (0) | 0 (0) | 0 (0) | 6 (0) | 0 (0) | 0 (0) | 0 (0) |

identified with common testing techniques such as unit and integration testing. Security concerns, such as inadvertent removal of bounds checks, can be identified using common static analysis tools like Coverity [26].

## 3. Evaluation

To evaluate CARVE, we selected four software packages implementing network protocols that vary in size, feature density, and complexity. Debloating network protocol implementations is a challenging and realistic problem due to their complex interactions and adherence to strict specifications. The software packages we selected are:

- libmodbus v3.1.4, an industrial protocol library.
- Bftpd v4.9, an FTP server utility program.
- libcurl v7.61.0, a data transfer utility library.
- mongoose v.6.8, an embedded web server library.

For each package, we manually created feature mappings at various levels of granularity, as well as three different debloating configurations (additional details are included in Appendix A). These configurations were selected to reflect reasonable real-world use cases corresponding to aggressive, moderate, and conservative debloating scenarios. We define these levels in the following common language:

- Conservative (C): Some peripheral features in the package are targeted for debloating.
- Moderate (M): Some peripheral features and some core features are targeted for debloating.
- Aggressive (A): All debloatable features except for a small set of core features are targeted for debloating.

In each scenario, we debloated the package with CARVE and built the resulting variant using the default package build configuration and GCC v7.3.0. All software packages and their debloated variants were built on the same platform, with configurations kept constant for each build. We tested all variants using a combination of developer-provided test scripts and custom testing scripts to ensure that they were correct with respect to kept features, and that inputs invoking debloated features did not cause the variant to crash.

### 3.1. Security Improvements

The most compelling security benefit of debloating software features is the potential to eliminate software vulnerabilities contained within these features. To demonstrate CARVE's potential for vulnerability elimination, we searched the CVE database [19] for known vulnerabilities affecting our selected software packages. There were no listed CVEs for libmodbus and Bftpd, however there are six known vulnerabilities affecting libcurl and 17 affecting mongoose. We analyzed the debloated versions of our selected packages to determine how many vulnerabilities were actually eliminated in our scenarios. Our results are shown in column two of Table 1.

Another commonly cited security benefit of debloating compiled software is increased resistance against code re-use attacks. Code re-use attacks redirect execution to existing instructions in memory to cause a malicious effect. In gadget-based code reuse attack methods such as ROP, JOP, and COP [15, 16, 17], execution is redirected to an ordered chain of short instruction sequences (gadgets) present in the program to construct a malicious payload without injecting code. Debloating unnecessary software features eliminates the gadgets found in the code implementing these features. Measuring the security benefit of debloating against gadget-based code reuse attacks is complicated because code-removing debloaters like CARVE, TRIMMER, and CHISEL may introduce new gadgets as a side effect of feature elimination [22]. We measure CARVE's effectiveness against gadget-based attacks using metrics proposed by Brown and Pande [22], special purpose gadget availability and gadget set expressivity. Our results are shown in the grouped columns of Table 1.

In general, CARVE is effective at reducing expressivity of gadget sets with respect to three different levels of expressivity [20, 21]. In seven of the twelve scenarios, debloating reduced the overall gadget set expressivity. However,

debloating had no effect in three scenarios and negative effects in two scenarios. CARVE was also generally successful at removing special purpose gadgets, achieving the highest reduction for `libcurl`. As was the case with gadget set expressivity, negative side effects were observed in two scenarios. The negative side effects observed in our results are consistent with the negative side effects observed in CHISEL [22] and TRIMMER [4]. Since CARVE is easy to reconfigure and debloating is fast, it is well suited for mitigating these negative side effects using iterative debloating [22].

### 3.2. Code Size Reduction

CARVE is also capable of significantly reducing the size of a software package. The average LOC, function count, and binary size reduction achieved at each level aggressiveness is displayed in Table 2. Interestingly, we observed slightly higher binary reduction than anticipated, which we attribute to improved performance by downstream compiler optimizations operating on simplified debloated code.

In some cases, debloating also eliminates dependencies on external libraries. This lowers runtime memory consumption as these libraries are no longer loaded during execution. Eliminating these libraries may also have positive security impacts if they may contain code and gadgets useful in a code reuse exploit. We did not observe a reduction in required libraries for `libmodbus`, `Bftpd`, and `mongoose`. While conservatively debloating `libcurl` only reduced the number of required libraries from 42 to 41, we observed a *two-thirds* reduction in required libraries for moderate and aggressive variants of `libcurl` (28 and 29 eliminated respectively).

Table 2: Average Code Size Reduction.

| Level of Aggressiveness | LOC Reduction | Function Count Reduction | Binary Size Reduction |
|---|---|---|---|
| Conservative (C) | 10.7% | 11.1% | 10.8% |
| Moderate (M) | 16.1% | 13.7% | 18.5% |
| Aggressive (A) | 31.3% | 27.1% | 33.0% |

### 3.3. Overhead Costs

CARVE's decoupled design minimizes per-debloat overhead. Debloating is fast and efficient, taking less than 5 seconds for each scenario. Debloating also scales effectively with increasing code size and density of feature mappings.

The vast majority of cost associated with this technique occurs during feature mapping. The amount of effort required to create and maintain feature mappings depends on a number of variables, including the user's familiarity with the software, feature density, and complexity of the target software. In practice, we found that creating the initial feature mapping was not cost-prohibitive in situations where the user has limited familiarity with the source code. Software engineering practices favor modularization and separation of concerns within software packages, which has the effect of reducing the number of feature mappings required to produce a sound mapping. Producing a mapping for our benchmark packages required 61.5 mappings per thousand LOC (m/KLOC) for `libmodbus`, 24.2 m/KLOC for `Bftpd`, 13.4 m/KLOC for `mongoose`, and 4.9 m/KLOC for `libcurl`. Additionally, these mappings did not require a significant number of complex code-replacing mappings; `Bftpd` and `mongoose` required no such mappings, `libmodbus` requires 17, and `libcurl` required 20. This data suggests that the one-time costs of creating feature mappings are reasonable.

## 4. Discussion

### 4.1. Comparison to Existing Techniques

Aggressively debloating simple programs (e.g. Linux CoreUtils) with CHISEL and TOSS results in average code size reductions of 86% and 43.8% respectively [2, 3]. Aggressively debloating more complex network-based software (e.g. cUrl) with TRIMMER reduces code size by 21% on average [4]. CARVE outperforms TRIMMER on similar benchmarks, and performs comparably to CHISEL and TOSS when accounting for benchmark selection. CARVE also performs comparably to CHISEL with respect to CVE elimination [3]. Finally, CARVE reduces gadget counts and introduces new gadgets at rates comparable to CHISEL and TRIMMER [22, 4].

### 4.2. Limitations

CARVE operates solely on source code and cannot be used to debloat binaries directly; it is not a suitable for closed-source or legacy software. While CARVE is technology independent and can be used on any software that is text-based and supports comments, at this time it implements only one language-specific debloating module: C/C++.

### 4.3. Future Work

In future versions of CARVE, we plan to create new language-specific debloating modules for other common programming languages such as Java and Javascript. We also plan to develop static analysis tools to assist the user in quickly producing sound feature mappings. Such tools would analyze mappings created by the user in real time to suggest new mappings and identify potential soundness issues associated with individual debloating operations.

## 5. Conclusion

In this paper, we introduced CARVE, a simple and powerful software debloating technique that overcomes the shortcomings of existing approaches. We demonstrated that effective debloating can be accomplished without the need for advanced software analysis and in a manner that preserves desirable software properties like interoperability and specification compliance. The results of our evaluation show that CARVE is well-suited for the challenging problem of debloating network protocols, improving the security and performance of four protocol implementations. Across 12 scenarios, CARVE eliminated vulnerabilities, reduced the utility of code reuse gadgets, and reduced overall code size on par with existing approaches using a simple technique that is accessible to the average user.

# A. Appendix

The following tables detail which features were mapped for each benchmark, and which were debloated in each scenario. For brevity, the complete list of fine-grained debloatable features in the package are condensed into the categories in the leftmost column. Features removed by CARVE are marked with an X in that scenario's column.

Table A.1: Debloated features per scenario for `libmodbus`.

| Debloatable Feature | Conservative Scenario | Moderate Scenario | Aggressive Scenario |
|---|---|---|---|
| RTU Read Operations | X | | X |
| RTU Write Operations | X | | X |
| RTU Raw Operations | X | | X |
| TCP (IPv4) Read Operations | | X | X |
| TCP (IPv4) Write Operations | | X | X |
| TCP (IPv4) Raw Operations | | X | X |
| TCP (IPv4/6) Read Operations | | X | |
| TCP (IPv4/6) Write Operations | | X | X |
| TCP (IPv4/6) Raw Operations | | X | X |

Table A.2: Debloated features per scenario for `Bftpd`.

| Debloatable Feature | Conservative Scenario | Moderate Scenario | Aggressive Scenario |
|---|---|---|---|
| Admin Commands | X | X | X |
| Read Commands | | | |
| Write Commands | | | X |
| Directory Commands | | X | X |
| Server Config Commands | | X | X |
| Miscellaneous Commands | | X | X |
| Server Info Commands | | | X |

Table A.3: Debloated features per scenario for `libcurl`.

| Debloatable Feature | Conservative Scenario | Moderate Scenario | Aggressive Scenario |
|---|---|---|---|
| Uncommon API Elements | X | X | X |
| HTTP | | | X |
| HTTPS | | | X |
| RTSP | X | X | X |
| FTP Read Commands | | | X |
| FTP Write Commands | | X | X |
| FTPS | | X | X |
| Telnet | X | X | X |
| LDAP | | X | X |
| TFTP | X | X | |
| IMAP | | X | X |
| SMB | | X | X |
| SMTP | | X | X |
| POP3 | | X | X |
| RTMP | X | X | X |
| File | X | X | X |
| Gopher | X | X | X |
| Dict | X | X | X |
| SCP | X | X | X |
| SFTP | | X | X |

Table A.4: Debloated features per scenario for `mongoose`.

| Debloatable Feature | Conservative Scenario | Moderate Scenario | Aggressive Scenario |
|---|---|---|---|
| Threads API | | | X |
| Broadcast API | X | | |
| IPV6 | X | | X |
| COAP | | | X |
| DNS | | X | X |
| HTTP Server | X | | |
| HTP CGI | X | | |
| Websocket | X | | |
| HTTP Client | | | X |
| Websocket | | | |
| MQTT | | | X |
| SNTP | X | | X |